 \documentclass[lettersize,journal]{IEEEtran}
\usepackage{amsfonts}
\usepackage{graphicx}
\usepackage{color}
\usepackage{amsmath,amsfonts,amssymb,amsthm,epsfig,epstopdf,url,array}
\usepackage{url,textcomp}
\usepackage{authblk}
\usepackage{cite}
\usepackage{verbatim}
\usepackage{subfigure}
\usepackage{algorithm}
\usepackage{algorithmic}
\usepackage{bbm}
\usepackage{amsmath,amsfonts}
\usepackage{algorithmic}
\usepackage{algorithm}
\usepackage{array}
\usepackage[caption=false,font=normalsize,labelfont=sf,textfont=sf]{subfig}
\usepackage{textcomp}
\usepackage{stfloats}
\usepackage{url}
\usepackage{verbatim}
\usepackage{graphicx}
\usepackage{cite}

\def\BibTeX{{\rm B\kern-.05em{\sc i\kern-.025em b}\kern-.08em
    T\kern-.1667em\lower.7ex\hbox{E}\kern-.125emX}}
\makeatletter

\newcommand{\Rmnum}[1]{\expandafter\@slowromancap\romannumeral #1@}
\makeatother
\begin{document}
\title{Deep Reinforcement Learning Empowered Rate Selection of XP-HARQ}
\author{Da~Wu,
        Jiahui~Feng,
        Zheng~Shi,
        Hongjiang Lei,
        Guanghua Yang,
        and Shaodan Ma
\thanks{This work was supported in part by National Natural Science Foundation of China under Grants 62171200, 62171201, 61971080, and 62261160650, in part by Chongqing Key Laboratory of Mobile Communications Technology under Grant cqupt-mct-202204, in part by Guangdong Basic and Applied Basic Research Foundation under Grant 2023A1515010900, in part by Zhuhai Basic and Applied Basic Research Foundation under Grant ZH22017003210050PWC, in part by the Major Talent Program of Guangdong Provincial under Grant 2019QN01S103, amd in part by the Science and Technology Development Fund, Macau SAR under Grants 0087/2022/AFJ and SKL-IOTSC(UM)-2021-2023. (\emph{Corresponding Author: Zheng Shi.})}
\thanks{Da~Wu, Jiahui~Feng, Zheng Shi, and Guanghua Yang are with the School of Intelligent Systems Science and Engineering, Jinan University, Zhuhai 519070, China (e-mails: 0x8a@stu2021.jnu.edu.cn; jiahui@stu2020.jnu.edu.cn; zhengshi@jnu.edu.cn; ghyang@jnu.edu.cn).}
\thanks{H. Lei is with Chongqing Key Lab of Mobile Communications Technology \& Chongqing University of Posts and Telecommunications, Chongqing 400065, China (e-mail: leihj@cqupt.edu.cn).}
\thanks{Shaodan Ma is the State Key Laboratory of Internet of Things for Smart City, University of Macau, Macau, China (e-mail: shaodanma@um.edu.mo).}
        }
\maketitle
\begin{abstract}
The complex transmission mechanism of cross-packet hybrid automatic repeat request (XP-HARQ) hinders its optimal system design. To overcome this difficulty, this letter attempts to use the deep reinforcement learning (DRL) to solve the rate selection problem of XP-HARQ over correlated fading channels. In particular, the long term average throughput (LTAT) is maximized by properly choosing the incremental information rate for each HARQ round on the basis of the outdated channel state information (CSI) available at the transmitter. The rate selection problem is first converted into a Markov decision process (MDP), which is then solved by capitalizing on the algorithm of deep deterministic policy gradient (DDPG) with prioritized experience replay. The simulation results finally corroborate the superiority of the proposed XP-HARQ scheme over the conventional HARQ with incremental redundancy (HARQ-IR) and the XP-HARQ with only statistical CSI.


\end{abstract}
\IEEEpeerreviewmaketitle
\hyphenation{HARQ}
\section{Introduction}\label{sec:int}

Hybrid automatic repeat request (HARQ) is one of the key technologies that is capable of offering reliable transmissions. However, this benefit is essentially reaped at the price of large transmission delay, 
which is unfavorable for fulfilling the ultra-reliable and low-latency communications (URLLC). 
To resolve such a dilemma, there is a urgent need to develop a flexible HARQ transmission mechanism that could be reconfigurable to meet diverse URLLC requirements. In this letter, we focus on the cross-packet HARQ (XP-HARQ) that is an evolutionary version of HARQ with high spectral efficiency, albeit at the price of high complexity \cite{jabi2017adaptive,jabi2017boost,jabi2018amc}. Unlike the conventional HARQ schemes, new information bits are introduced in retransmissions such that surplus wireless resources are substantially exploited. Hence, it is unnecessary to wait for the end of the retransmissions of the current message before the delivery of the next message especially under benign channel conditions. As a consequence, the spectral efficiency of HARQ is boosted, meanwhile the average transmission delay is reduced.

Recently, the investigations on the XP-HARQ scheme are still in their fancy. Several efforts have been made to accurately evaluate and optimally design XP-HARQ schemes. In \cite{jabi2017adaptive}, Mohammed Jabi {\emph {et al.}} examined the long term average throughput (LTAT) of XP-HARQ, with which the throughput improvement gained by XP-HARQ was verified. In \cite{jabi2017boost}, a two-layer coding scheme was developed to implement XP-HARQ to guarantee the inputs of the encoder with the same length, where puncturing and mixing operations were leveraged. The puncturing rates were then optimized with dynamic programming in \cite{jabi2017boost}. The adaptive modulation and coding scheme was further introduced to boost the LTAT of XP-HARQ in \cite{jabi2018amc}. In \cite{shi2019effective}, the effective capacity of XP-HARQ was analyzed for buffer-limited XP-HARQ. However, the performance metrics of XP-HARQ in \cite{jabi2017adaptive,jabi2017boost,jabi2018amc,shi2019effective} were obtained by conducting Monte-Carlo simulations and lacked insightful analysis. To fill this vacancy, the most fundamental performance metric, namely, outage probability, was derived in closed-form for XP-HARQ over independent Rayleigh fading channels in \cite{9768121}, with which full time diversity of XP-HARQ was proved. However, even under such a simple channel model, the outage analysis is too complex to further assist the optimal design of XP-HARQ, not to mention under more complicated fading channels.

To address the above issue, we resort to the data-driven deep reinforcement learning (DRL) for the optimal design of XP-HARQ over correlated fading channels. It should be noticed that only a few works attempted to devise the conventional HARQ schemes using the DRL methods. Particularly, in \cite{9376717}, a DRL enabled user scheduling policy was designed to minimize the age of information (AoI) for HARQ systems. In \cite{9217354}, a deep deterministic policy gradient (DDPG) algorithm was leveraged to maximize the throughput via optimizing the incremental redundancy bits. Unfortunately, the extension of the DRL methods to general HARQ schemes has never been reported. This letter maximizes the LTAT via adaptive rate selection by considering outdated channel state information (CSI). The optimization problem is firstly formulated as a problem of Markov decision process (MDP). By taking into account the continuous state and action spaces, the problem is then solved by using DDPG with prioritized experience replay. By conducting Monte Carlo simulations, the proposed XP-HARQ scheme is proved to be superior to the conventional HARQ with incremental redundancy (HARQ-IR) and the XP-HARQ with only statistical CSI. Furthermore, it is found that the time correlation among fading channels does not lead to a significant impact upon the LTAT of the proposed XP-HARQ scheme.

The rest of this letter is outlined as follows. Section \ref{sec:sys mod} introduces the system model. Section \ref{sec:rein} develops a DRL empowered rate selection algorithm for XP-HARQ.
The simulated results are presented 
in Section \ref{sec:eva}. Section \ref{sec:con} finally concludes this letter.
\begin{figure}[!t]
    \centering
    \includegraphics[width=8cm]{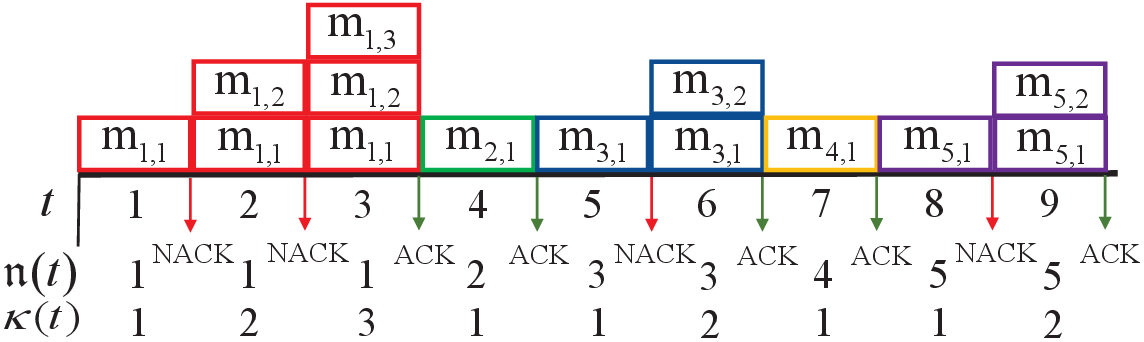}
    \caption{An example of the XP-HARQ scheme with $K=3$.}
    \label{Fig.XP-HARQ model} 
\end{figure}

\section{System Model}\label{sec:sys mod}
This letter considers a point-to-point communication system, in which XP-HARQ is adopted to enable the retransmissions of the message. To start, this section delineates the system model, including the XP-HARQ transmission mechanism, the channel model, performance metrics, and the rate selection problem.

\subsection{XP-HARQ}

As shown in Fig. \ref{Fig.XP-HARQ model}, an example is used to illustrate the transmission mechanism of the XP-HARQ. To avoid network congestion in unfavorable propagation environment, the number of transmissions of XP-HARQ is limited up to $K$. For notational simplicity, let $\frak{n}(t) \in \mathbb Z^{+}$ and $\kappa(t) \in [1,K]$ be the functions that map the time slot $t$ to the current HARQ cycle and the current transmission round, respectively. In the initial transmission round of the $\frak{n}(t)$-th HARQ cycle, the message ${\rm m}_{\frak{n}(t),1}$ is encoded as a codeword ${\bf x}_{\frak{n}(t),1}$ with a transmission rate $R_1$. The received signal ${\bf y}_{\frak{n}(t),1}$ reads as
\begin{equation}
    {\bf y}_{\frak{n}(t),1} = \sqrt{P_1} h_{\frak{n}(t),1} {\bf x}_{\frak{n}(t),1} + {\bf n}_{\frak{n}(t),1},
\end{equation}
where $h_{\frak{n}(t),1}$ denotes the channel coefficient of the first round of the $\frak{n}(t)$-th HARQ cycle with ${\mathbb E}(|h_{\frak{n}(t),1}|^2)=1$, ${\bf n}_{\frak{n}(t),1}$ stands for the complex additive Gaussian noise (AWGN) having zero mean and a variance of $\sigma^2$, and $P_1$ is the average transmit power in the initial HARQ round. If ${\bf x}_{\frak{n}(t),1}$ is successfully decoded, a positive acknowledgement (ACK) will be sent back to confirm the successful reception of ${\rm m}_{\frak{n}(t),1}$ and the next HARQ cycle with index $t+1$ will be triggered immediately. Otherwise, a negative acknowledgement (NACK) will be fed back to initiate the retransmissions. According to the coding strategy of XP-HARQ \cite{jabi2017adaptive}, as opposed to the conventional HARQ-IR that only redundant information bits are retransmitted, new information bits are introduced in the retransmissions by XP-HARQ to substantially exploited wireless resources. Accordingly, prior to the $\kappa(t)$-th transmission of the $\frak{n}(t)$-th HARQ cycle, the previously failed messages ${\rm m}_{\frak{n}(t),1},\cdots,{\rm m}_{\frak{n}(t),\kappa(t)-1}$ are combined with the currently received message ${\rm m}_{\frak{n}(t),\kappa(t)}$ to form a longer  message ${{\rm m}_{\frak{n}(t),[\kappa(t)]}}$. The concatenated message ${{\rm m}_{\frak{n}(t),[\kappa(t)]}}$ is encoded as a codeword ${\bf x}_{\frak{n}(t),\kappa(t)}$ with a nominal transmission rate $\sum\nolimits_{\kappa=1}^{\kappa(t)} R_{\kappa}\triangleq R_{\kappa(t)}^{\Sigma}$, where the increment of the transmission rate, i.e., $R_{\kappa}$, originates from the new information bits involved in the $\kappa$-th transmission. Therefore, the signal $\mathbf{y}_{\frak{n}(t),\kappa(t)}$ received in the ${\kappa(t)}$-th round of the current XP-HARQ cycle is written as
\begin{equation}\label{eqn:receive signal 1}
\mathbf{y}_{\frak{n}(t),\kappa(t)}=\sqrt{P_{\kappa(t)}} h_{\frak{n}(t),\kappa(t)} \mathbf{x}_{\frak{n}(t),\kappa(t)}+\mathbf{n}_{\frak{n}(t),\kappa(t)},
\end{equation}
where $h_{\frak{n}(t),\kappa(t)}$, ${\bf n}_{\frak{n}(t),\kappa(t)}$, and $P_{\kappa(t)}$ follow the similar definitions as $h_{\frak{n}(t),1}$, ${\bf n}_{\frak{n}(t),1}$, and $P_1$, respectively, which are omitted here to save space. The messages ${\rm m}_{\frak{n}(t),1},\cdots,{\rm m}_{\frak{n}(t),\kappa(t)}$ are jointly decoded by using the observations ${y_{1}}, \cdots,{y_{\kappa(t)}}$. The current XP-HARQ cycle stops and the next process begins once the receiver succeeds in reconstructing all the previously delivered messages or the maximum number of HARQ transmission attempts ${\kappa(t)}$ is used. Interested readers are referred to \cite{7878518} for more details of the encoding/decoding implementation of XP-HARQ.




\subsection{Channel Model}
This letter considers time-correlated Rayleigh flat-fading channels, where the channel keeps constant during each codeword transmission slot and changes time-dependently across consecutive transmission slots. We define $t$ as the index of the time slot in the sequel. 
For notational simplicity, we use the notation $\hbar _{t} $ to represent $ h_{\frak{n}(t),\kappa(t)} $. 
 As a commonly used time-correlated channel model that takes place in the environment of low-to-medium mobility, $ \hbar_{t} $ is modeled according to a first-order Gauss-Markov process as \cite{5710995}, i.e., 
\begin{equation}\label{eqn:state transition}
    {\hbar}_{t}=\rho {\hbar}_{t-1}+\sqrt{1-\rho^{2}} {w}_t,
\end{equation}

where ${\rho}$ is the correlation coefficient between ${\hbar}_{t}$ and ${\hbar}_{t-1}$, ${w}_t \sim \mathcal{C N}\left({0}, \sigma^{2} \right)$ denotes the channel discrepancy and is independent of ${\hbar_{t-1}}$. In order to account for the impact of channel aging, the outdated channel state ${\hbar}_{t-1}$ is sent back to the transmitter.

\subsection{Performance Metrics}
\subsubsection{Outage Probability}
The outage probability is an essential performance metric for evaluating the system reliability. The outage probability of XP-HARQ is the probability of the event that the accumulated mutual information in each HARQ round is below the transmission rate. More specifically, the outage probability of XP-HARQ after $K$ HARQ rounds is given by \cite{jabi2017adaptive}
\begin{equation}\label{eq:out}
    f_K = \Pr\left(I_1<R_1,I_2 < R_2^\Sigma,\cdots,I_K < R_K^\Sigma\right),
\end{equation}
where ${I_k} = \sum_{l=1}^k{\log _2}(1 + |{\hbar_l}{|^2}{{{P_l}} {/{{\sigma ^2}}}})$ stands for the accumulated mutual information until the $l$-th transmission.

\subsubsection{Long Term Average Throughput}
The long term average throughput (LTAT) is a frequently used performance metric to evaluate the expected throughput of HARQ systems \cite{Caire2001}. 
The LTAT of XP-HARQ system is defined as \cite{jabi2017adaptive}
\begin{equation}\label{eqn:LTAT}
    \begin{aligned}
    \eta_K
    &=\lim_{T\to \infty} \frac{\mathcal R(T)}{T} =\frac{\sum_{k=1}^K R_k\left(f_{k-1}-f_K\right)}{1+\sum_{k=1}^{K-1} f_k},
    \end{aligned}
\end{equation}
where ${\mathcal R(t)}$ refers to the total number of successfully received information bits till time $t$, and the second equality in \eqref{eqn:LTAT} is derived in \cite{Caire2001,jabi2017adaptive} by capitalizing on the renewal theory if only the statistical CSI is available at the transmitter. 
\subsection{Maximization of LTAT}
This paper aims to maximize the LTAT through optimal rate selection if only the aged channel state information (CSI) is available at the transmitter. The optimization problem of the transmission rates can be formulated as
\begin{equation}\label{eq:ltat_max}
 \begin{array}{*{20}{c l}}
{\mathop {\max }\limits_{{R_1}, \cdots ,{R_K}} }& \eta_K\\
{{\rm{s}}{\rm{.t}}{\rm{.}}}&{0 \le {R_k} \le \bar R,\,k\in [1,K]},
\end{array}
\end{equation}
where the transmission rate $\{R_k,\,k\in [1,K]\}$ is upper bounded by $\bar R$ to avoid frequent outages because of the limited resources. 
However, due to the time correlation among fading channels in \eqref{eqn:state transition} and the involved outage definition in \eqref{eq:out}, it is hardly possible to get the explicit outage expression. Hence, it is unlikely to solve the LTAT maximization problem in \eqref{eq:ltat_max} with the conventional optimization tools. To overcome this difficulty, we recourse to the deep reinforcement learning (DRL) for the optimal solution of the transmission rate.

\section{DRL Empowered Rate Selection}\label{sec:rein}
Due to the rapid change of time-varying fading channels, it results in a prohibitively high system overhead to acquire the instantaneous CSI. Therefore, we assume that only the outdated and statistical CSIs are available at the transmitter, including the channel state of the previous slot $\hbar _{t-1}$ and the correlation coefficient $\rho$. Moreover, the transmission rate of the current transmission round for XP-HARQ is determined by the transmission status (success or failure), rates, and channel states in the previous transmission rounds. Towards this end, the proposed optimization problem is transformed into a Markov decision process (MDP), which can be solved with DRL methods.

\subsection{Problem Reformulation and MDP} 


By using the definition of the LTAT and replacing the limit operation with the expectation (the time average converges to the ensemble average for ergodic processes), the original problem \eqref{eq:ltat_max} can be reformulated as
\begin{equation}\label{eq:ltat_max_ref}
\begin{array}{*{20}{c l}}
{\mathop {\max }\limits_{{R{(t)}}} }& \mathbb E \left(\frac{\mathcal R(T)}{T} \right)=\mathbb E \left(\frac{1}{T}{{\sum\nolimits_{t = 1}^{T} {{\mathcal R_{\frak{n}(t),\kappa(t)}}} }}\right)\\
{{\rm{s}}{\rm{.t}}{\rm{.}}}&{0 \le {R{(t)}} \le \bar R},
\end{array}
\end{equation}
where the expectation is taken over the randomness of the channel states, ${R{(t)}}$ is the effective transmission rate for the new information bits in the time slot $t$, 
${{\mathcal R_{\frak{n}(t),\kappa(t)}}}$ denotes the effective transmission rate for the successfully received information bits after $\kappa(t)$ rounds during the $\frak{n}(t)$-th HARQ cycle. According to the Shannon theory, the successful decoding occurs if and only if the transmission rate is less than the channel capacity. Therefore, ${{\mathcal R_{\frak{n}(t),\kappa(t)}}}$ can be obtained as
\begin{equation}\label{eqn:Throughput}
    {{\mathcal R_{\frak{n}(t),\kappa(t)}}} = \left\{ {\begin{array}{*{20}{c}}
        {R^\Sigma_{\kappa(t)} ,}&{\quad I_{\kappa(t)}  \ge R^\Sigma_{\kappa(t)}}\\
        {0,}&{{\rm{else}}}
        \end{array}} \right..
\end{equation}

With the problem reformulation of \eqref{eqn:Throughput}, the adaptive rate selection scheme can be modeled as an MDP, which can be solved by leveraging reinforcement learning (RL) method. The MDP essentially comprises four elements, including environment $\mathcal E$, state space $\mathcal S$, action space $\mathcal A$, and reward space $\mathcal R$. More specifically, at each time step $t$, the process is in state $s_t\in \mathcal S$. According to the current state, the agent makes a decision to choose an action $a_t \in \mathcal A$. After taking the action $a_t$, the next state $s_{t+1}$ is observed along with a reward $r_{t}\in \mathcal R$ received from the environment $\mathcal E$. By mapping the optimal rate selection of XP-HARQ as an MDP, the states, actions, and rewards are designed as follows.
\subsubsection{State $s_t$}
To capture the channel aging effect, the historical channel state ${h}_{t}$ is considered into the observation of environment. Moreover, the decoding status of XP-HARQ essentially depends on the accumulated mutual information and rate. Accordingly, the state $s_t$ is a vector consisting of the previously accumulated transmission rate and mutual information intended for the $\frak{n}(t)$-th XP-HARQ, and the aged channel state ${h}_{t-1}$, namely 
    \begin{equation}\label{eqn:state}
    {s_t} \buildrel \Delta \over = \left\{ {\begin{array}{*{20}{c}}
{\left( {R_{\kappa (t - 1)}^\Sigma ,{I_{\kappa (t - 1)}},{\hbar_{t - 1}}} \right),}&\frak{n}(t-1)=\frak{n}(t)\\
{\left( {0,0,{\hbar_{t - 1}}} \right),}&{\rm else}
\end{array}} \right.,
    \end{equation}
    wherein the accumulated transmission rate and mutual information for the current HARQ cycle are zero if a new HARQ cycle is initiated, i.e., $\frak{n}(t-1)\ne \frak{n}(t)$.
\subsubsection{Action $a_t$} 
The action is defined as the effective transmission rate for the new information bits in the next HARQ round, i.e.,
    \begin{equation}\label{eqn:action}
    {a}_{t} \triangleq {R{(t)}}.
    \end{equation}
\subsubsection{Reward $r_t$} 
    The reward function can be defined as the effective transmission rate of the successfully received information bits for the current HARQ cycle $\frak{n}(t)$, i.e.,
    \begin{equation}\label{eqn:reward}
    {r_t} = r({s_t},{a_t},{s_{t+1}}) \triangleq  {{\mathcal R_{\frak{n}(t),\kappa(t)}}}.
    \end{equation}
    By noticing the continuous space of the states and actions, the MDP problem can be solved with the DRL, which combines the reinforcement learning and deep neural networks to learn the policy. The details are deferred to the next subsection.

\subsection{DRL Empowered Rate Selection }
\begin{figure}[htbp]
    \centering
    \includegraphics[width=3.5in]{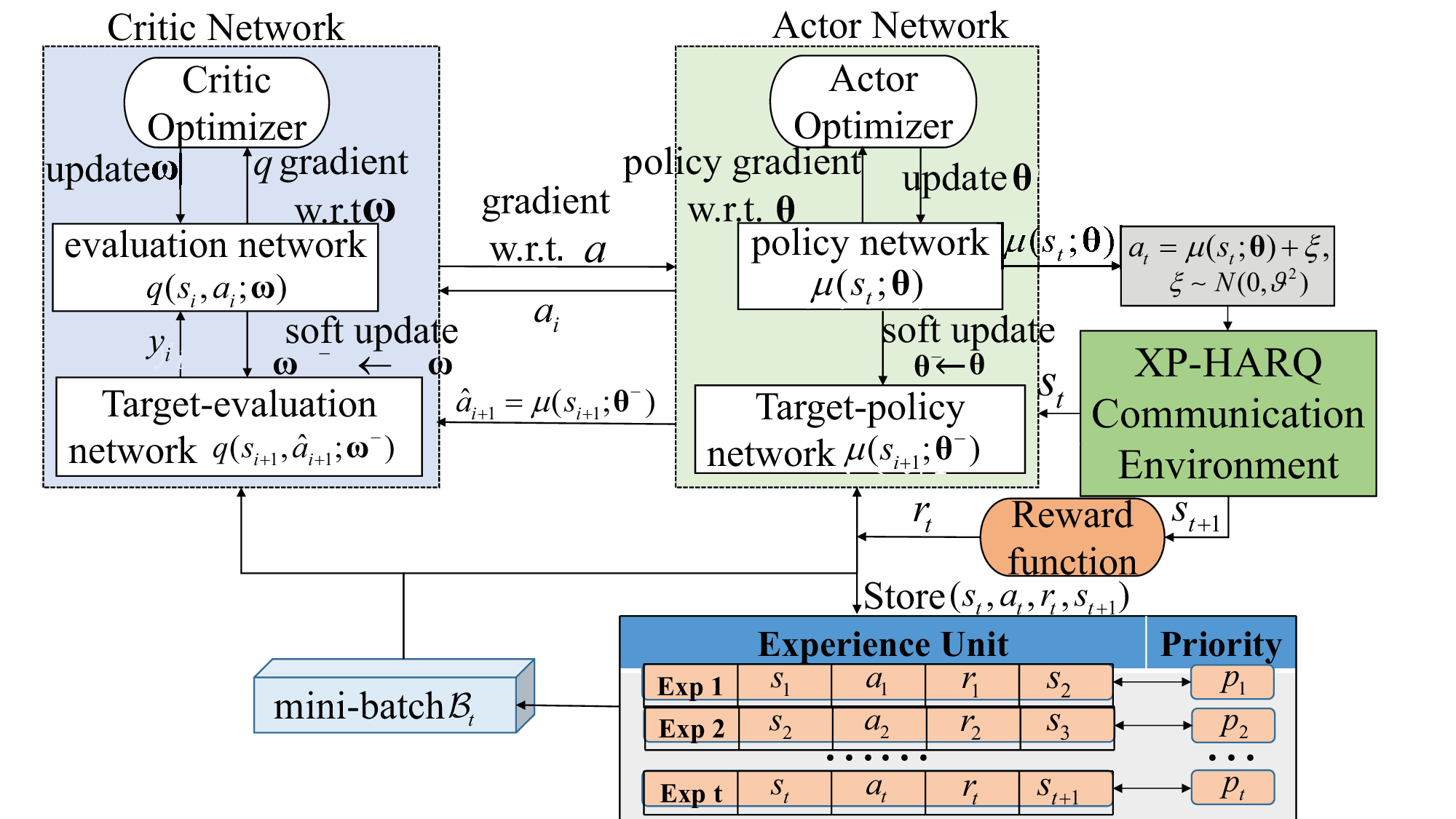}
    \caption{The DDPG network for Rate Selection of XP-HARQ.}
    \label{Fig.DDPG_framework} 
\end{figure}

A DRL based rate selection scheme is proposed for the LTAT maximization of the XP-HARQ. By considering the continuous state and action spaces, a deep deterministic policy gradient (DDPG) with prioritized experience replay will be applied to develop the rate selection framework, as shown in Fig. \ref{Fig.DDPG_framework}. This framework consists of four neural networks, i.e., two policy networks (also termed as the actor network, i.e., $\mu(s_t;\boldsymbol{\theta})$ and $\mu(s_{t+1};\boldsymbol{\theta}^-)$) and two evaluation networks (also termed as the critic network, i.e., $Q(s_t,a_t;\boldsymbol{\omega})$ and $Q(s_{t+1},\hat a_{t+1};\boldsymbol{\omega}^-)$), wherein the target-evaluation and target-policy networks are used to calculate the temporal-difference (TD) target to address the overestimation issue, and these neural networks are parameterized by $\boldsymbol{\theta}$, $\boldsymbol{\theta}^-$, $\boldsymbol{\omega}$, and $\boldsymbol{\omega}^-$. In addition, for the stability and fast convergence, a prioritized experience reply memory pool $\mathcal {M}$ is adopted to collect the agent's experience tuple $e_t = (s_i,a_i,r_i,s_{i+1})$ at each time $t$. At each time step, the four neural networks will be updated with a mini-batch of experience samples $\mathcal B_t$ that are drawn from $\mathcal {M}$ according to the priority of the playback experience, that is, $e_t \sim  \mathcal P (\mathcal M) $ for $\forall e_t \in \mathcal B_t$, where $\mathcal P $ is the probability function defined in \eqref{eqn:prob_def}.
In what follows, priority experience playback mechanism and the training processes of the four neural networks are described in detail.

\subsubsection{Prioritized Experience Replay}
In contrast with the uniform random experience replay, the prioritized experience replay is capable of accelerating the learning process and enhancing the training stability \cite{schaul2015prioritized}. According to the prioritized sampling strategy, the sampling probability $p_i$ of the tuple ${e_i = (s_i , a_i , r_i , s_{i+1})}$ is proportional to the absolute value of TD error $\delta_i$, i.e., 
\begin{equation}\label{eqn:prob_def}
    p_i  \propto |\delta_i| + \epsilon,
\end{equation}
where $ \epsilon$ is a positive constant to avoid a zero sampling probability, ${\delta _i} = Q({s_i},{a_i};{\boldsymbol{\omega }}) - {r_i} - \gamma Q({s_{i + 1}},{{\hat a}_{i + 1}};{{\boldsymbol{\omega }}^ - })$ denotes the TD error, and $\gamma$ is the discount factor.


\subsubsection{Evaluation Network}
The evaluation network aims to approximate the actual state-action function $Q_\pi(s,a)$ with a neural network parameterized by ${\boldsymbol{\omega}}$. The network parameters ${\boldsymbol{\omega}}$ can be updated with the TD algorithm. More specifically, the loss function is defined as the weighted squared TD error averaged over the sampled mini-batch $\mathcal B_t$, i.e.,
\begin{equation}\label{eqn:evaLoss}
L({\boldsymbol{\omega}} ) = \frac{1}{{2|\mathcal{B}_t|}}{\sum\limits_{e_i \in \mathcal B_t}{w_i \delta _i^2}},
\end{equation}
where $|\mathcal{B}_t|$ represents the batch size and the importance-sampling weight $w_i$ is used to eliminate the bias introduced by prioritized sampling and ensure the same learning rate of all samples. According to \cite{schaul2015prioritized}, $w_i$ is given by
\begin{equation}\label{eqn:weight_def}
     w_i \propto {\left(|\mathcal{B}_t| p_i\right)^{-\beta}},
\end{equation}
which $\beta \in [0,1]$ is a hyperparameter that controls the extent of the correction. Then, the gradient descent algorithm is leveraged to update the network parameters  $\boldsymbol{\omega }$ as
\begin{equation}\label{eqn:omega_up}
   \boldsymbol{\omega }_{\rm new} \leftarrow  \boldsymbol{\omega }_{\rm now} - \alpha {\nabla _{\boldsymbol{\omega }}}L({\boldsymbol{\omega }_{\rm now}}),
\end{equation}
where ${\nabla _{\boldsymbol{\omega }}}L({\boldsymbol{\omega }}) = \frac{1}{{|{{\cal B}_t}|}}\sum\nolimits_{{e_i} \in {{\cal B}_t}} {{w_i}{\delta _i}{\nabla _{\bf{\boldsymbol{\omega} }}}Q({s_i},{a_i};{\bf{\omega }})} $ refers to the gradient of the loss function with respect to (w.r.t.) $\boldsymbol{\omega }$, and $\alpha$ is the learning rate.
\subsubsection{Policy Network}
The policy network $\mu(s_t;\boldsymbol{\theta})$ aims to learn action policy by mapping the states to the specific actions. Since the action-value function $Q_\pi(s,a)$ can evaluate the score of the current action policy, the performance objective for $\mu(s_t;\boldsymbol{\theta})$ can be defined as \cite{silver2014deterministic}
\begin{equation}\label{eqn:evaLoss}
J({\boldsymbol{\theta }}) = \frac{1}{{|{{\cal B}_t}|}}\sum\limits_{{e_i} \in {{\cal B}_t}} {Q({s_i},\mu ({s_i};{\boldsymbol{\theta }});{\boldsymbol{\omega }_{\rm now}})}.
\end{equation}
To learn the best policy, the parameters of the policy network can be optimized through the maximization of $J({\boldsymbol{\theta }})$. Accordingly, the gradient ascend method is used to update $\boldsymbol{\theta }$, i.e.,
\begin{equation}\label{eqn:theta_up}
   \boldsymbol{\theta }_{\rm new} \leftarrow  \boldsymbol{\theta }_{\rm now} + \upsilon {\nabla _{\boldsymbol{\theta }}}J({\boldsymbol{\theta }_{\rm now}}),
\end{equation}
where $\upsilon $ is the learning rate, and using chain rule yields ${\nabla _{\boldsymbol{\theta }}}J({\boldsymbol{\theta }}) = \frac{1}{{|{{\cal B}_t}|}}\sum\nolimits_{{e_i} \in {\mathcal B_t}} {{\nabla _{\boldsymbol{\theta }}}\mu ({s_i};{\boldsymbol{\theta }}){\nabla _a}Q({s_i},{{\hat a}_i};{{\boldsymbol{\omega }}_{{\rm{now}}}})} $.

\subsubsection{Target Evaluation/Policy Networks}
To further improve the stability, the soft update strategy is applied to update the parameters of the target networks, i.e., $\boldsymbol{\omega ^ - }$ and ${\boldsymbol {\theta ^ - }}$. More specifically, with the new parameters $\boldsymbol{\omega }_{\rm new}$ and $\boldsymbol{\theta }_{\rm new}$ given by \eqref{eqn:omega_up} and \eqref{eqn:theta_up}, respectively, the parameters of the two target networks will be updated as
\begin{equation}
    {\boldsymbol {\omega }_{\rm new} ^ -} \leftarrow \tau {\boldsymbol {\omega}_{\rm new}}  + (1 - \tau) {\boldsymbol {\omega }_{\rm now} ^ -},
\end{equation}
\begin{equation}
    {\boldsymbol {\theta }_{\rm new} ^ -} \leftarrow \tau {\boldsymbol {\theta}_{\rm new}}  + (1 - \tau) {\boldsymbol {\theta}_{\rm now} ^ - },
\end{equation}
where the hyperparameter $\tau\ll 1$.
\section{Simulations and Discussions}\label{sec:eva}
In this section, simulated results are presented for verifications and discussions. For illustration, the system parameters are set as $\sigma^{2} = 1$, ${\rho}=0.4$, and ${\bar R}=10 $~bps/Hz unless otherwise specified. Besides, we assume equal power allocation for XP-HARQ, i.e., $P_{1}=\cdots=P_{K}$, and the average transmit signal-to-noise ratio (SNR) is defined as $P_{1} / \sigma^{2}= \cdots= P_{K} / \sigma^{2} \triangleq { \textsc{snr}}$. 
To deploy the DDPG, both the actor and critic networks consist of one input layer, three hidden layers, and one output layer. The number of the neurons in the three hidden layers are 100, 50, and 30 neurons, respectively. The three hidden layers of both networks use ``{\emph{ReLu}}'' activation functions. The output layer of the actor network invokes ``{\emph{sigmoid}}'' activation function to restrict the transmission rate within $\bar R$, while the critical network does not leverage any activation function in the output layer. 
Both the actor and critical networks capitalize on the adaptive moment estimation (Adam) optimizer to update the network parameters, and the learning rates are set to $\upsilon = \alpha = 0.001$. Furthermore, we assume that the number of epochs in the training state is 100, the  number of time slots in each epoch is 6000, the size of the prioritized replay buffer is $|\mathcal M|=20000$, the mini-batch size is $|\mathcal B_t|=512$. In addition, we assume that the weight of the soft update $\tau=0.01$, the discount factor $\gamma=0.9$, the extent of the correction $\beta=0.5$, and the noise variance of the behavior policy $\vartheta^2 = 0.2$.



\begin{figure}[htbp]
    \centering
    \includegraphics[width=7cm]{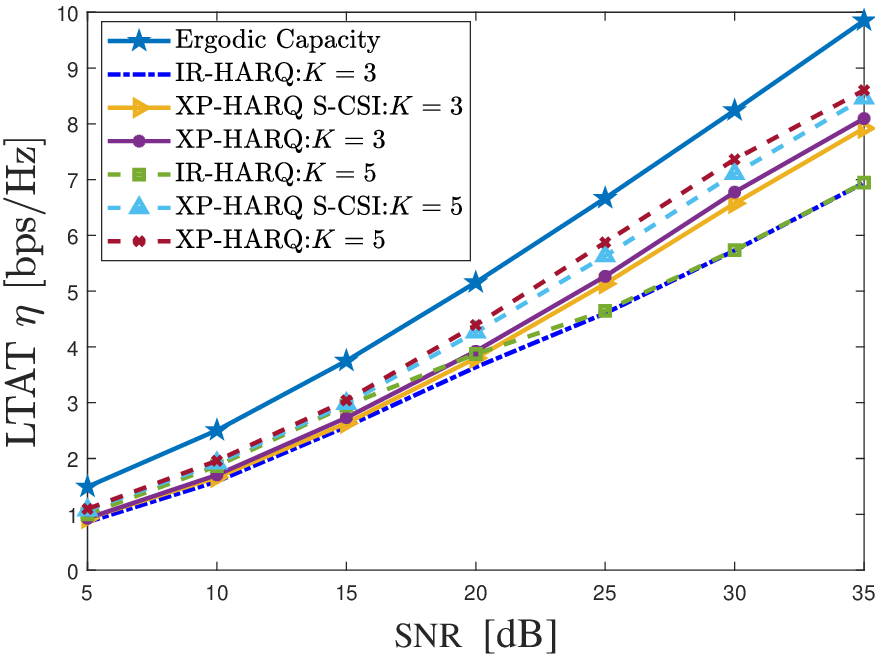}
    \caption{The comparison of the LTAT for different HARQ schemes.}
    \label{Fig.throughputs} 
\end{figure}

Fig. \ref{Fig.throughputs} depicts the LTAT performance of XP-HARQ versus of the average transmit SNR under different $K$. To exhibit the superiority of the proposed DRL-empowered rate selection scheme, two baseline HARQ schemes are used for comparison, including the conventional HARQ-IR \cite{7959548} 
 and the XP-HARQ with only statistical CSI (labeled as ``S-CSI'' in the figure) \cite{9768121}. The results of XP-HARQ with S-CSI can be regarded as the worst performance limit of our proposed scheme. In the meantime, the ergodic capacity is incorporated for benchmarking purpose or as design guidelines. It is shown in Fig. \ref{Fig.throughputs} that the XP-HARQ scheme performs much better than the HARQ-IR scheme. For example, by fixing ${ \textsc{snr}} = 35$~dB and $K=5$, the XP-HARQ scheme achieves a higher LTAT than the HARQ-IR scheme by around 1.65~bps/Hz. It is also seen from Fig. \ref{Fig.throughputs} that the proposed XP-HARQ scheme with outdated CSI surpasses the XP-HARQ scheme with statistical CSI by around 0.15~bps/Hz. Moreover, as the maximum number of transmissions $K$ increases from 3 to 5, a remarkable performance gain can be attained by both XP-HARQ schemes with the outdated CSI and the statistical CSI, whereas the HARQ-IR scheme achieves a negligible LTAT enhancement particularly at high SNR. This advantage of XP-HARQ attributes to new information bits introduced in retransmissions. Moreover, this merit also brings about a reduced transmission delay.


%

%
%

\begin{figure}[htbp]
    \centering
    \includegraphics[width=7cm]{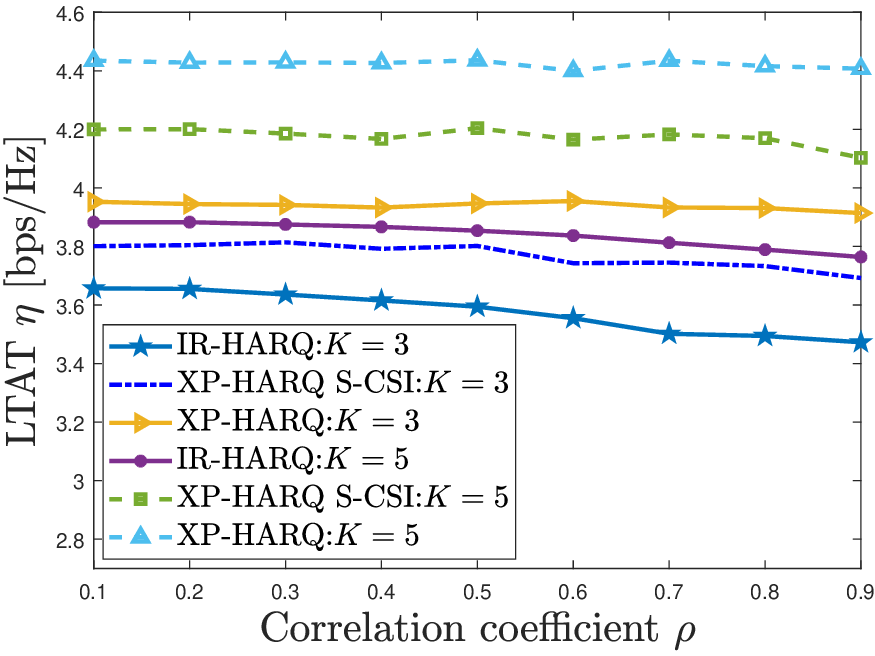}
    \caption{Impact of correlation coefficient $\rho$.}
    \label{Fig.Throughput_Rho} 
\end{figure}

Fig. \ref{Fig.Throughput_Rho} investigates the impact of the time correlation coefficient on the LTAT given a fixed ${ \textsc{snr}} = 20$~dB. Overall, it is not beyond our expectation that the time correlation has a detrimental effect on the LTAT. This is because more time diversity gain can be achieved from fading channels with a lower time correlation \cite{7959548}. Nevertheless, it is noteworthy that the superiority of the proposed XP-HARQ schemes essentially stems from utilizing the outdated CSI. Hence, a low channel correlation will result in less similarity of CSIs between two adjacent transmissions, which limits the time diversity gain from retransmissions. Accordingly, it can be seen from Fig. \ref{Fig.Throughput_Rho} that the LTAT curves slightly decrease with $\rho$.

%

\section{Conclusion}\label{sec:con}
Due to the lack of simple analytical results of the performance metrics of XP-HARQ, we applied the DRL to properly select the incremental information rate for XP-HARQ over correlated fading channels, without recourse to the traditional optimization tools. More specifically, the maximization of the LTAT was formulated as a problem of MDP, which can be solved by using the algorithm of DDPG with prioritized experience replay. To demonstrate the efficacy of the proposed XP-HARQ scheme, its LTAT performance was compared to the conventional HARQ-IR and the XP-HARQ with only statistical CSI through simulations. It was found that IR-HARQ is more aggressive than XP-HARQ when determining the initial rate. In the meantime, it was also found that the time correlation has a slightly negative impact on the LTAT of the proposed XP-HARQ scheme.

\bibliographystyle{IEEEtran}
\bibliography{References}
\end{document}